# Three-photon electron spin resonances

S. I. Atwood[1], V. V. Mkhitaryan[2], S. Dhileepkumar,[1] C. Nuibe,[1] S. Hosseinzadeh[1], H. Malissa[1,2], J. M. Lupton[1,2], and C. Boehme[1]

[1]Department of Physics and Astronomy, University of Utah, Salt Lake City, Utah 84112, USA

[2]Institut für Experimentelle und Angewandte Physik, Universität Regensburg, Universitätsstrasse 31, 93053 Regensburg, Germany

**Abstract:**

We report the observation of a three-photon resonant transition of charge-carrier spins in an organic light-emitting diode using electrically detected magnetic resonance (EDMR) spectroscopy at room temperature. Under strong magnetic-resonant drive (drive field $B_1$ ~ static magnetic field $B_0$), a $B_0$-field swept EDMR line emerges when $B_0$ is approximately threefold the one-photon resonance field. Ratios of drive-induced shifts of this line to those of two- and one-photon shifts agree with analytical expressions derived from the Floquet Hamiltonian and confirm the nature of these three-photon transitions, enabling access of spin physics to a hitherto inaccessible domain of quantum mechanics.

After almost a century of studies on multi-photon transitions [1-5], leading to groundbreaking applications for spectroscopy [6], there has recently been increasing interest in the study of multi-photon magnetic dipole transitions of spin states [7-10], motivated not only by easier experimental access through alternative spin-resonance detection schemes such as electrically detected magnetic resonance (EDMR), but also by the prospect that the understanding of and the access to highly non-equilibrium strong-drive quantum phenomena that relate to spin states have the potential to further enhance their utility as qubits, in particular for sensing applications [11], and allow for dressed-state coherence-protection schemes [12]. Multi-photon magnetic dipole transitions can be observed under conditions of nonlinear strong drive, where a linearly polarized driving field $\boldsymbol{B}_1$ is no longer well approximated by a circularly polarized field (rotary wave approximation), i.e., when the ratio of the magnitude of $B_1$ and the static magnetic field $B_0$, which causes Zeeman splitting, approaches unity [8,10]. Theoretical predictions of the existence and qualitative nature of two-level, monochromatic, two-photon magnetic dipole transitions have recently been confirmed through observations of EDMR line shifts from charge-carrier spin pairs in organic light emitting diodes (OLEDs) [10]. In contrast to mere resonance artefacts caused by higher excitation harmonics [7], the presence of two-photon transitions was verified experimentally by measuring the shifts of these resonance lines as a function of the drive-field amplitude $B_1$. In principle, with high enough $B_1$, any $n$-photon resonance may be observed using this approach. In practice, though, few experimental reports of $n$-photon magnetic dipole resonances with $n > 2$ from monochromatic, continuous wave (cw) excitation of two-level systems have been made [9,13-17]. Even fewer studies have quantitatively scrutinized potential multi-photon resonances using theoretical predictions [16,17], and, to our knowledge, none have demonstrated this for electron or nuclear spin transitions so far, as past studies of two- [8,10] and one-photon [8,18] spin-resonances did not allow for the identification of three-photon and higher-order processes, mainly due to the limitations of the strength of $B_1$ when inductively non-resonant radiofrequency (RF) coils were used for these experiments.

In this letter, we revisit the question of a spin-resonant three-photon transition and its behavior under strong drive, specifically, its drive-induced resonance shift, which allows for its unambiguous quantitative verification. Figure 1(a) illustrates an energy-level diagram of a three-photon absorptive transition, which occurs through intermediate virtual states (shown as solid gray lines) between two spin eigenstates $|\downarrow\rangle$ and $|\uparrow\rangle$, which the electromagnetic radiation shifts energetically away from the Zeeman separation $\hbar\gamma B_0$, where $\hbar$ is the reduced Planck constant and $\gamma$ is the gyromagnetic ratio [10].

The experimental data presented in this study were obtained with an approach conceptually similar to recent studies focusing on other strong-drive magnetic-resonance effects that represent the dynamics of charge-carrier spin-pair permutation symmetry [8-10,18], i.e. OLED EDMR. Technically, the experiments presented here differ from these earlier studies significantly, as much larger $B_1$ magnitudes and many more accumulated datasets were needed to observe the much weaker three-photon signatures reliably. We conducted cw EDMR spectroscopy using spin-dependent recombination currents in organic bipolar injection devices (essentially OLEDs) in which the π-conjugated polymer 'Super-Yellow' (SY) poly(1,4-phenylenevinylene) (SY-PPV) was used as the active layer. SY-PPV is a commercial organic semiconductor that allowed us to prepare the large number of OLED samples needed to generate a much larger data pool than previous studies that used custom-made deuterated polymer materials [8,10,18]. Figure 1(b) displays five examples of $B_1$-modulated (20 Hz modulation frequency), lock-in detected SY-PPV EDMR spectra, recorded within a magnetic field ($B_0$) range of ±20 mT, with 100 MHz RF incident drive, for an increasing magnitude of $B_1$ between ~1 mT and ~4 mT. All data discussed in this article were recorded at room temperature. The spectra in Figure 1(b) verify the strong drive conditions through a variety of qualitative effects that appear with increasing $B_1$ and that have been discussed in the literature: the influence of power broadening [7] on the one-photon lines; the inversion of the one-photon peaks due to spin-collectivity [18]; the two-photon resonance [8,10]; the Bloch-Siegert shift of the one-photon signal [8,9]; and the non-Bloch-Siegert-type shift of the two-photon resonance [10]. There is also a feature at 0 mT, which relates to the so-called oscillating magnetic field effect and arises due to the dc magnetoresistance induced by $B_1$ [19]. In addition to all these spectral features, at $B_0 \approx \pm 10$ mT the data of Figure 1(b) also reveal a pair of additional resonance peaks that are the focus of the present study. These additional lines cannot be attributed to trivial one-photon transitions caused by amplifier anharmonicities [8,10], as they display a shift toward $B_0 = 0$ mT with increasing $B_1$ when compared to the vertical dashed lines, which indicate an unperturbed three-photon resonance at $3\omega/\gamma = 10.705$ mT, where $\omega$ is the angular frequency of the driving RF magnetic field. Thus, the quantitative consistency of this shift with predictions from Floquet theory, as discussed in the following, is a crucial prerequisite for the corroboration of this signal as a three-photon electron spin resonance signature.

For the experiments discussed here, the use of a commercially available, fully hydrogenated active OLED layer comes at the cost of needing much higher field amplitudes $B_1$ than would be necessary to observe strong-drive phenomena in perdeuterated compounds [8]. To generate these fields, the OLEDs were placed inside an RF coil that was part of a resonant RF impedance-matching circuit developed to maximize the power transfer from the RF amplifier to the coil (see Supplemental Material [20]). All measurements presented in this study were conducted with 100 MHz cw RF field under square-wave modulation at either 20 Hz, to ensure a fully steady state, or at 1 kHz, yielding comparable EDMR spectra with better signal-to-noise ratio, as discussed in the Supplemental Material [20]. The sweep rate for $B_0$ was chosen relative to the modulation frequency, the time constant of the lock-in amplifier, and the data transfer rate from the

Gaussmeter, in order to determine $B_0$ for the raw data collected and avoid passage effects and lock-in artefacts. To account for the absolute and relative inaccuracies of the Gaussmeter, the measured $B_0$ scale was corrected by using the one-photon EDMR resonance lines (at small $B_1$) as absolute magnetic field standards [10], reducing the systematic uncertainty of $B_0$ to ~150 µT at 1 kHz modulation and to ~20 µT at 20 Hz modulation.

All data collected for this study (see Supplemental Material [20]) were obtained from a total of 12 SY-PPV OLEDs, each measured separately. The tank circuit was tuned before each measurement series, so that each data set had a different conversion factor between the square root of applied power and $B_1$. For most experiments, power dependencies were measured from the lowest to the highest power with a monotonic increase, after the tank circuit and the OLED devices were allowed to reach the steady state at the operating point over more than 4 minutes each time the power was changed. When the RF power increased, the radiation-induced, non-spin-dependent electric currents also increased, and therefore necessitated device bias adjustments to maintain a constant current (10 µA) for all measurements. After reaching the highest input power, which was chosen to be the point where Ohmic heating of the RF coil led to a detuning of the tank circuit and caused $B_1$ to drop even with increasing power, we reversed the sequence and proceeded to lower power. Most devices degraded to a point of failure (an abrupt change of the IV curve to a non-diodic behavior) within 24 hours, a shorter lifetime than typically observed under weak-drive conditions. The intense radiation-induced electric currents might have contributed to such faster degradation. For all measurements, the RF excitation current was monitored through a series resistor, as discussed in the Supplemental Material [20].

Altogether, 1959 EMDR spectra were recorded from all OLEDs. For the analysis of the experimental data, the spectra were subjected to a first-order (linear) baseline correction to account for slow sample drift, and the resonance center fields were determined by averaging the three maxima (or minima for inverted one-photon resonances at high $B_1$). This method deviates from the procedure reported in Ref. [10], which estimated the line centers by fitting a second-order polynomial to the resonance peaks. In most cases, the two methods yielded the same result, within error, yet the newly introduced method was much faster and no less accurate for the significantly larger data pool. Altogether, 3943 resonance line centers were obtained from the spectra. The uncertainty in the resonance line centers was dominated by the uncertainty of the $B_0$ scale, while the uncertainty in the amplitude resulted from the baseline noise. Averages of data were made only under truly identical experimental conditions, i.e., comparable input power and either ascending or descending $B_1$, but not combinations of both. A weighted average was calculated if all values were within two standard deviations of all the others; otherwise, the mean and standard deviation of the mean ($\sigma/\sqrt{N}$) were calculated.

Figure 2 shows the values of the peak centers $B_c$ for one-, two-, and three-photon resonances as a function of the measured current in the tank circuit, $I_{RF}$, for spectra obtained from one OLED at 20 Hz modulation. The peak centers as a function of $I_{RF}$ for all 12 devices are shown in the Supplemental Material [20]. The peak centers decrease monotonically with increasing $I_{RF}$, which is proportional to $B_1$ by a factor that changes when the resonance frequency of the tank circuit changes, e.g., when resistive heating of the RF coil becomes significant [20]. While this complication prevents a quantitative comparison of the data plotted in Fig. 3 with theoretical predictions, the monotonic relationship between the observed line shifts and $I_{RF}$ clearly shows that the observed resonance lines cannot be attributed to amplifier harmonics or other artefacts that would display a different shift behavior. As can been seen in the raw data [20], the shift in the one-photon peak extrema displays a discontinuity at or near the onset of the one-photon peak inversion. This effect has been discussed previously in the context of strong magnetic-resonance drive [10] and is caused by the onset of spin collectivity [8,18]. Given this effect and the uncertainty of $B_1$, a rigorous scrutiny of the shift values obtained requires theoretical predictions that are independent of $B_1$. We recently derived such predictions, which require only comparison against one- and two-photon resonance line shifts recorded at the same time as the three-photon signal, i.e., under identical $B_1$ conditions [10]. Analytical expressions for $n$-photon resonance peak centers, $B_{\gamma_n}$, are

$$B_{\gamma_1} \simeq \frac{\omega}{\gamma} - \frac{\gamma B_1^2 \sin^2\theta}{16\omega}, \qquad n = 1,$$

$$B_{\gamma_n} \simeq n\frac{\omega}{\gamma} - \frac{\gamma B_1^2 \sin^2\theta}{4\omega}\frac{n}{n^2-1}, \qquad n > 1,$$

(1)

with $\theta$ the angle between $\boldsymbol{B}_1$ and $\boldsymbol{B}_0$, i.e. $\theta = 90°$ when the two vectors are perpendicular, as in standard magnetic resonance conditions. (Please note that Eq. (1) differs from Eq. (15) in Ref. [10] because $\theta$ is defined differently here. In Ref. [10], $\theta$ was defined as the deviation of $\boldsymbol{B}_1$ from $\boldsymbol{B}_1 \perp \boldsymbol{B}_0$.) Rearranging these equations and defining the driving-field-induced resonance shift $\Delta B_{\gamma_n} \equiv n\,\omega/\gamma - B_{\gamma_n}$, we find that

$$\frac{\Delta B_{\gamma_n}}{\Delta B_{\gamma_1}} = \frac{4n}{n^2-1}$$

$$\frac{\Delta B_{\gamma_n}}{\Delta B_{\gamma_m}} = \frac{n}{m}\frac{m^2-1}{n^2-1}$$

(2)

for any $n, m > 1$. The ratio of the two- to one-photon shift from Eq. (2) is $\Delta B_{\gamma_2}/\Delta B_{\gamma_1} = 8/3 = 2.67$, which we recently verified experimentally [10]. Since we are testing here for the three-photon resonance, we need to verify the ratios $\Delta B_{\gamma_3}/\Delta B_{\gamma_2} = 9/16 = 0.5625$ and $\Delta B_{\gamma_3}/\Delta B_{\gamma_1} = 3/2 = 1.5$. An experimental confirmation of these ratios only requires accurate estimates of the shifts of two different $n$-photon resonances, irrespective of $B_1$. Although Eq. (1) cannot be used to estimate the expected shift without knowing $B_1$, it can be used to estimate $B_1$ through the measured shift $\Delta B_{\gamma_n}$ and the angle $\theta$.

Figure 3 shows the EDMR peak shifts under strong drive for (a) $\Delta B_{\gamma_3}$ vs. $\Delta B_{\gamma_2}$, (b) $\Delta B_{\gamma_3}$ vs. $\Delta B_{\gamma_1}$, and (c) $\Delta B_{\gamma_2}$ vs. $\Delta B_{\gamma_1}$ obtained for all samples and recorded with $B_1 \perp B_0$. The solid lines pass through the origin and indeed have the slopes predicted by Eq. (2). Each plot shows the same range in the $x$- and $y$-axes to emphasize the different slopes. Generally, the shift ratios across all OLED samples show excellent agreement with the predicted ratios. In the data shown in panels Figure 3(b, c), which include the one-photon resonances, the shifts deviate from the predicted slope at lower shift values. This deviation is attributed to the discontinuity in the one-photon shifts close to the peak inversion (due to the onset of the spin-Dicke effect [18]) as discussed above, and are ultimately due to the absence of a discernable center of the resonance line. For this reason, the strong-drive limit is defined in the context of Figure 3 to include only data at higher RF than the discontinuity. The delineation of the data according to this definition is seen more clearly in the individual plots of peak centers vs. RF current in the Supplemental Material [20]. The shift ratios obtained are plotted separately for each individual sample in the Supplemental Material [20], together with a further plot of the data in Figure 3 on normalized scales. As expected, the shift ratios obtained under 1 kHz modulation are the same as those recorded under 20 Hz modulation. Some offsets from the predicted ratios in individual data sets may arise from the fact that the peak extrema are used to approximate the resonance line centers, a reasonable approximation as long as adjacent broad resonance lines do not overlap.

Following the predictions for multi-photon resonance line shifts under strong drive derived from the Floquet Hamiltonian in Ref. [10], we realize that the effective amplitude $u$ of the three-photon resonance under linearly polarized RF excitation is given by

$$u = \frac{\gamma^3 B_1^3}{32\omega^2}\left|\sin\theta - \frac{9}{8}\sin^3\theta\right|, \qquad (3)$$

an expression with a strong dependence of the angle $\theta$ between $B_1$ and $B_0$. This relation has minima at $\theta = 0°$ and $70.5°$ and maxima at $\theta = 33°$ and $90°$. While the sample holder was not designed to enable an adjustment of $\theta$ to arbitrary values, it was possible to change $\theta$ from $90°$ to $69(1)°$ and to conduct measurements on three different devices to ensure reproducibility of the results [20]. Figure 4 shows two experimental EDMR spectra obtained from the same OLED at the highest RF power. The black spectrum was recorded first with $B_1 \perp B_0$, while the gray spectrum was recorded shortly afterward with $\theta = 69°$. We note several qualitative differences between the two spectra. The inverted one-photon resonance at $69°$ is not as strong as its counterpart at $90°$, while, conversely, the two-photon line is more intense. These characteristics are qualitatively consistent with the predictions from theory in Ref. [10]. Most significantly, the three-photon resonance is not discernible above the noise level for the non-orthogonal arrangement of $B_1$ and $B_0$. According to Eq. (3), the relative intensity of the three-photon resonance between 90° and 69° is 2.5. We can check these ratios through the statistics of the four spectra measured at $\theta = 90°$ (of which the black spectrum is one) and the six spectra recorded at $\theta = 69°$ (of which the gray spectrum is one). The mean three-photon resonance intensity of the four spectra at 90° is 0.0038±0.0007 arb. u., while the mean of the baseline noise at 69° is 0.0009±0.0003 arb. u. The ratio of the two is therefore 4±2, which, within the error, agrees with the theoretical prediction. Additional spectra, results from numerical simulations, and the values of the averaged peak centers as a function of the drive current $I_{RF}$ at $\theta = 69°$ for each of the three OLEDs utilized, as well as the shift ratios for $\Delta B_{\gamma_2}$ vs. $\Delta B_{\gamma_1}$, are given in the Supplemental Material [20]. Note that none of the peak centers from the angled data are shown in Figure 3 because they do not fall within the strong-drive limit defined above for Figure 3.

In conclusion, we find that room temperature, cw, low-field, strong-drive EDMR spectroscopy of charge carrier spin states in the π-conjugated polymer SY-PPV confirms the existence of three-photon magnetic-dipole transitions through the characteristic shifts of the resonance lines with drive strength and magnetic field orientation as predicted by theory [8,10]. The result of this study, confirming the existence of three-photon resonant spin transitions as well as elucidating their nature, opens up pathways to a variety of previously inaccessible spin effects and their potential applications, including the use of three-photon transitions for spin-qubit initialization and manipulation, as well as strong magnetic-drive induced protection of spin coherence [8].

**Acknowledgements**

This work was supported by the U.S. Department of Energy, Office of Basic Energy Sciences, Division of Materials Sciences and Engineering under Award #DE-SC0000909. H. M. and V. V. M. acknowledge support from the Deutsche Forschungsgemeinschaft (DFG, German Research Foundation) (project ID 314695032 – SFB 1277, subproject B03). S. D. and C. N. acknowledge support through the University of Utah's College of Science Student Research Initiative.

**Figures**

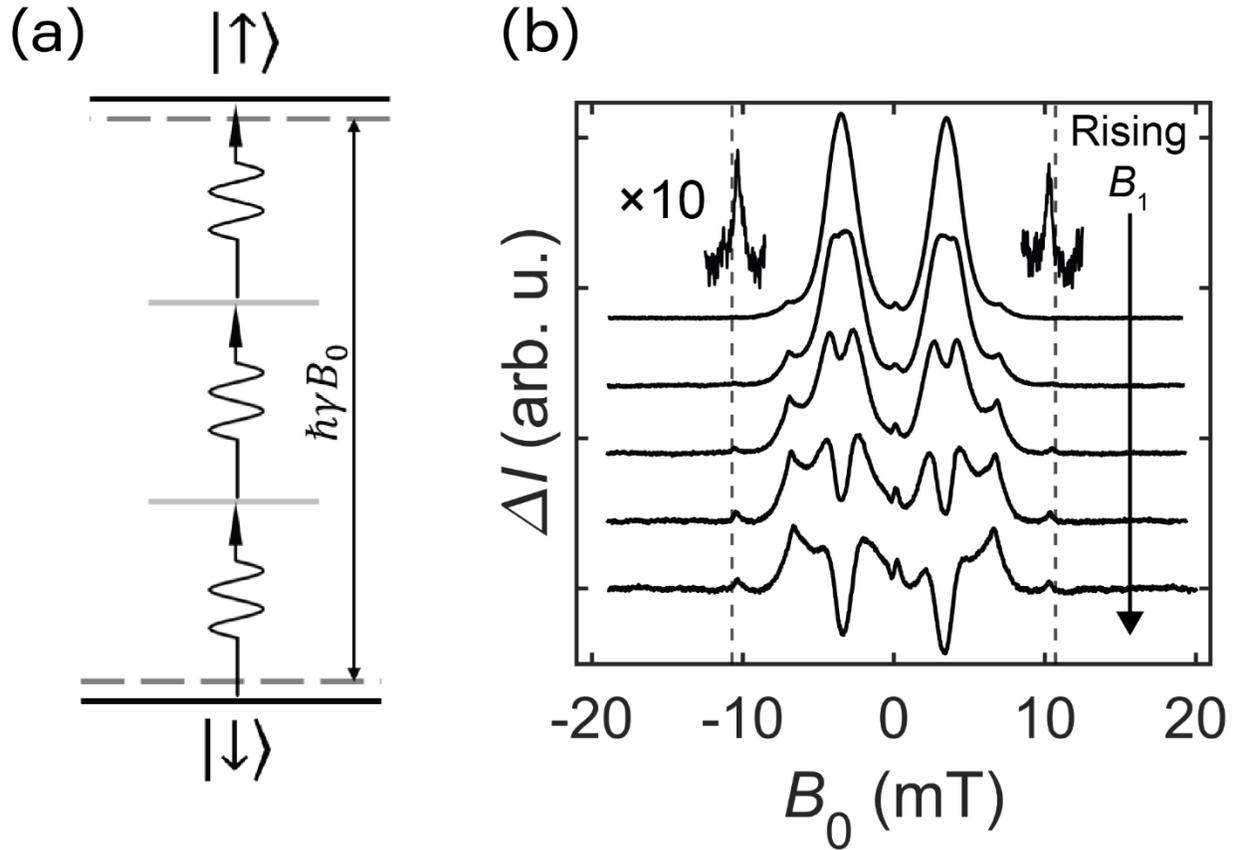

Figure 1. Electrically detected magnetic resonance (EDMR) spectra of an OLED at high drive-field strengths. (a) Energy-level diagram of a three-photon absorptive transition, which occurs through intermediate virtual states (solid gray lines) between two spin eigenstates $|\downarrow\rangle$ and $|\uparrow\rangle$, which the electromagnetic radiation shifts energetically away from the Zeeman separation $\hbar\gamma B_0$. (b) A selection of five strong-drive EDMR spectra from an OLED sample with SY-PPV recorded with RF-modulated (20 Hz), lock-in detected cw spectroscopy. The drive field amplitude increases from top to bottom. The magnified three-photon resonances belong to the bottom-most spectrum. The vertical dashed lines indicate an unshifted three-photon resonance. The one-photon resonance bifurcates and inverts due to the ac-Zeeman effect and the spin-Dicke effect [18]. The peak at zero field arises from the quasistatic magnetic-field effect [19].

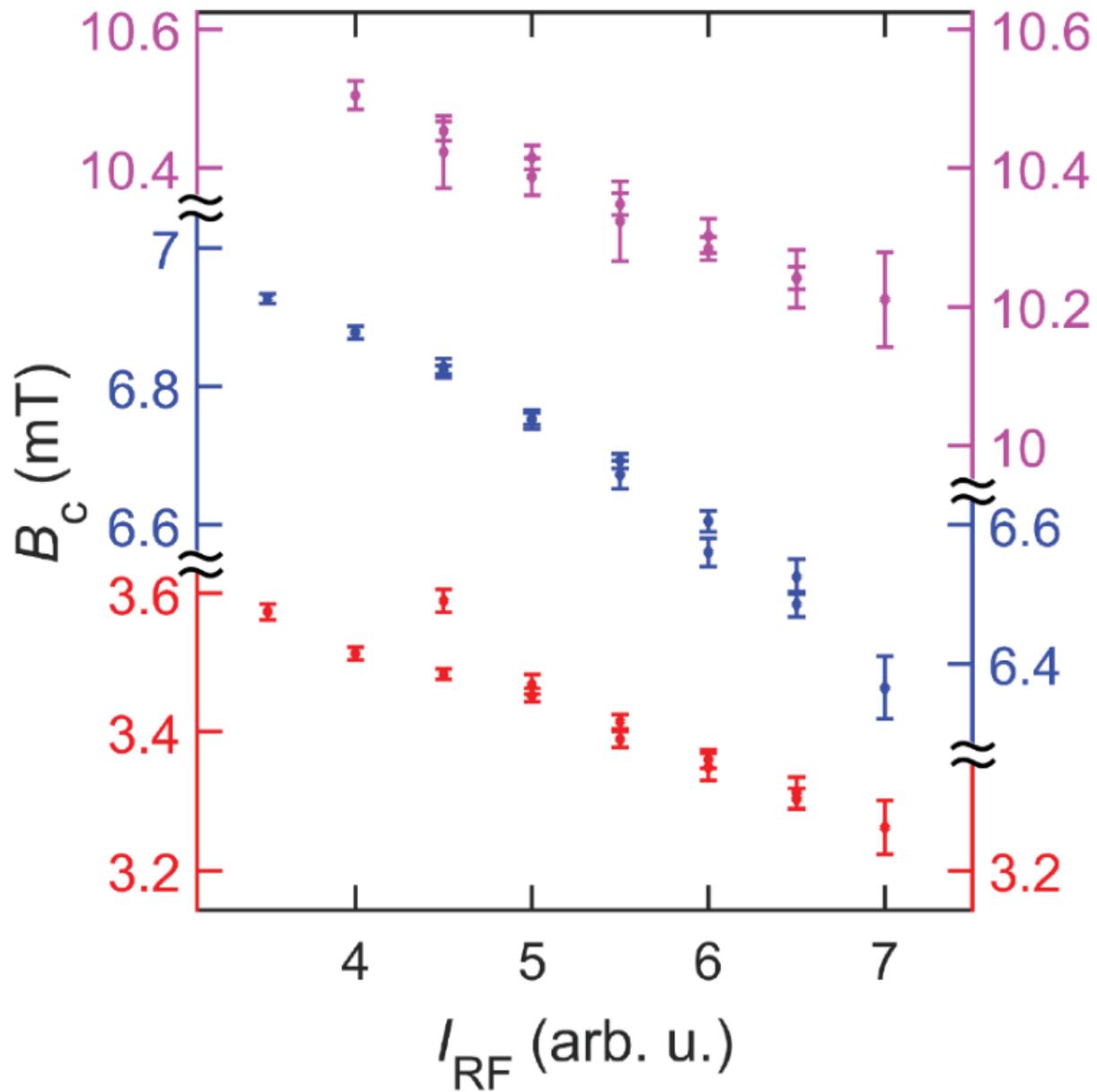

Figure 2. Averaged peak centers $B_c$ (through identification by the resonance line extrema) as a function of current in the RF circuit (which is proportional to $B_1$) for one-, two-, and three-photon resonances, identified by red, blue, and magenta, respectively. The right and left scales correspond to identical units, yet have different scale breaks to accommodate all data in the same plot.

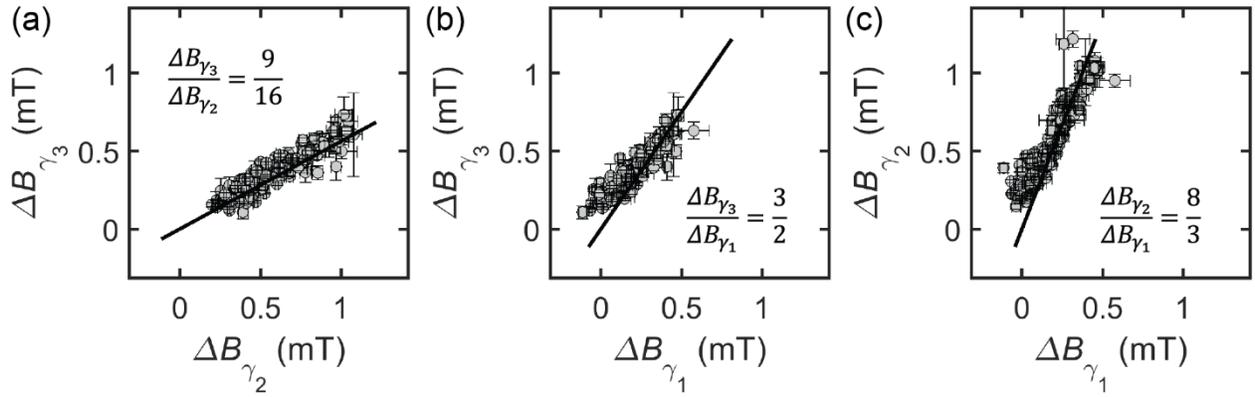

Figure 3. Plots of drive-induced resonance line shifts of (a) three- vs. two-photon, (b) three-vs. one-photon, and (c) two- vs. one-photon transitions of all measured data sets that fall within the strong-drive limit as defined in the main text. The solid lines pass through the origin and have slopes equal to the ratios stated in the panels. These ratios are predicted by Eq. (2).

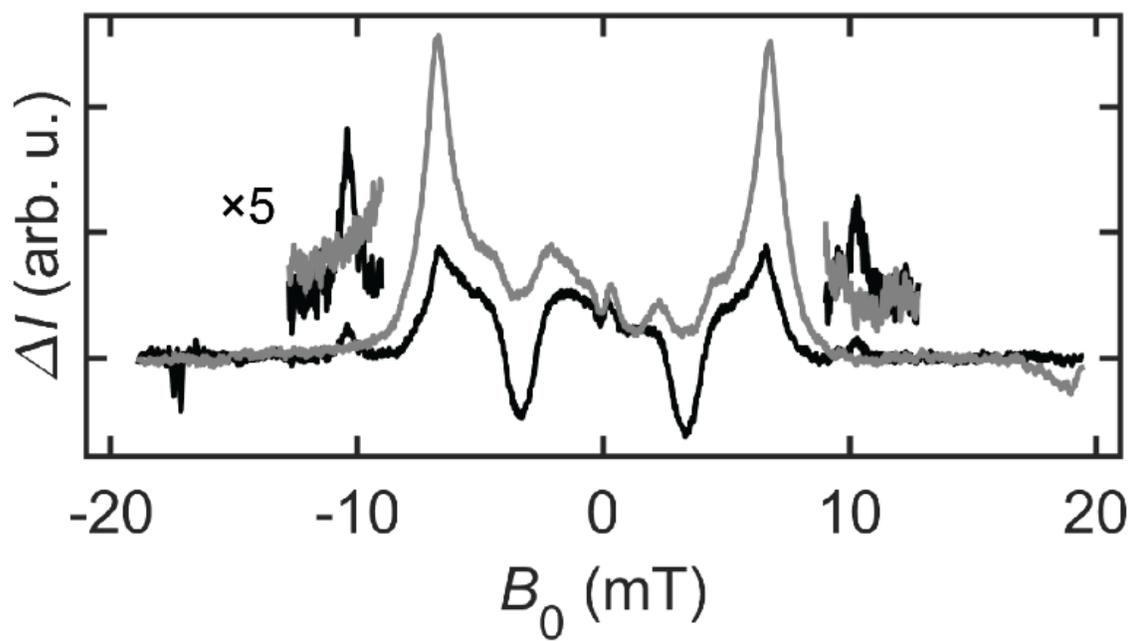

Figure 4. OLED EDMR spectra obtained at the same nominal $B_1$ strengths for $\theta = 69°$ (gray) and $\theta = 90°$ (black). The former data do not show any discernable three-photon resonances.

# References

1. M. Göppert-Mayer. *Ann. Phys.* **401**, 273 (1931). https://doi.org/10.1002/andp.19314010303

2. V. Hughes and L. Grabner. *Phys. Rev.* **79**, 314 (1950). https://doi.org/10.1103/PhysRev.79.314

3. L. Grabner and V. Hughes. *Phys. Rev.* **82**, 561 (1951). https://doi.org/10.1103/PhysRev.82.561.2

4. J. Brossel, B. Cagnac, and A. Kastler. *C. R. Acad. Sci.* **237**, 984 (1953).

5. P. Kusch. *Phys. Rev.* **93**, 1022 (1954). https://doi.org/10.1103/PhysRev.93.1022

6. H. Mahr. Two-Photon Absorption Spectroscopy. *Quantum Electronics: A Treatise*. Edited by H. Rabin & C.L. Tang. Academic Press, New York (1975). pp. 285-361.

7. S. Jamali, G. Joshi, H. Malissa, J. M. Lupton, and C. Boehme. *Nano Lett.* 17, 4648 (2017). https://doi.org/10.1021/acs.nanolett.7b01135

8. S. Jamali, V. V. Mkhitaryan, H. Malissa, A. Nahlawi, H. Popli, T. Grünbaum, S. Bange, S. Milster, D. M. Stoltzfus, A. E. Leung, et al., *Nat. Commun*. 12, 465 (2021). https://doi.org/10.1038/s41467-020-20148-6

9. J. P. Ashton and P. M. Lenahan, *Phys. Rev. B* 102, 020101 (2020). https://doi.org/10.1103/PhysRevB.102.020101

10. S. I. Atwood, S. Hosseinzadeh, V. V. Mkhitaryan, T. H. Tennahewa, H. Malissa, W. Jiang, T. A. Darwish, P. L. Burn, J. M. Lupton, C. Boehme. arXiv:2310.14180 (cond-mat.mes-hall) (2023). Under review at *Phys. Rev. B*. https://doi.org/10.48550/arXiv.2310.14180

11. C.L. Degen, F. Reinhard, and P. Cappellaro. *Rev. Mod. Phys.* **89**, 035002-1 (2017). https://doi.org/10.1103/RevModPhys.89.035002

12. K. C. Miao, J. P. Blanton, C. P. Anderson, A. Bourassa, A. L. Crook, G. Wolfowicz1, H. Abe, T. Ohshima, D. D. Awschalom. *Science* **369**, 1493 (2020). https://doi.org/10.1126/science.abc5186

13. J. Margerie and J. Brossel. *C. R. Acad. Sci.* 241, 373 (1955).

14. B. Clerjaud and A. Gelineau. *Phys. Rev. Lett.* **48**, 40 (1982). https://doi.org/10.1103/PhysRevLett.48.40

15. V. Morozov, O. Antzutkin, A. Koptyug, and A. Doktorov. *Mol. Phys.* 73, 517 (1991). https://doi.org/10.1080/00268979100101361

16. D. Fregenal, E. Horsdal-Pedersen, L. B. Madsen, M. Førre, J. P. Hansen, and V. N. Ostrovsky. *Phys. Rev. A* 69, 031401 (2004). https://doi.org/10.1103/PhysRevA.69.031401

17. Y. Sun, Y. Xu, and Z. Wang. *Opt. Laser Technol.* 145, 107488 (2022). https://doi.org/10.1016/j.optlastec.2021.107488

18. D. P. Waters, G. Joshi, M. Kavand, M. E. Limes, H. Malissa, P. L. Burn, J. M. Lupton, and C. Boehme, *Nat. Phys.* 11, 910 (2015). https://doi.org/10.1038/nphys3453

19. S. Milster, T. Grünbaum, S. Bange, S. Kurrmann, H. Kraus, D. M. Stoltzfus, A. E. Leung, T. A. Darwish, P. L. Burn, C. Boehme, and J. M. Lupton.

20. Supplemental Material.

SUPPLEMENTAL MATERIAL

Three-photon electron spin resonances

S. I. Atwood[1], V. V. Mkhitaryan[2], S. Dhileepkumar,[1] C. Nuibe,[1] S. Hosseinzadeh[1], H. Malissa[1,2], J. M. Lupton[1,2], and C. M. Boehme[1]

[1]Department of Physics and Astronomy, University of Utah, Salt Lake City, Utah 84112, USA

[2]Institut für Experimentelle und Angewandte Physik, Universität Regensburg, Universitätsstrasse 31, 93053 Regensburg, Germany

A) OLED, RF coil, and impedance-matching circuit

Figure S1(a) depicts an OLED placed inside an RF coil that was part of a resonant RF impedance-matching circuit. The coil itself was placed within a copper box for shielding against stray, circuit-induced RF fields, as discussed below in Section (B). A full view of an OLED pixel, deposited on an EDMR contact template, is shown in Figure S1(b), along with a photograph of an OLED pixel emitting under bias. The pixel consists of the polymer active layer, combined with electron- and hole-injecting layers [S1-S3], and fabricated as described previously [S3-S5], with SY-PPV as the active layer and $MoO_3$ as the hole injection layer [S1,S6].

For the experiments described in the main text, the control of the bipolar static magnetic fields and the data acquisition were carried out as described in Ref. [S1]. For the generation of the RF field, a Dax22000 Wavepond arbitrary waveform generator (AWG) and a Mini-Circuits ZHL-100W-251-S+ amplifier (20-450 MHz, 100 W) were used. As in Ref. [S1], a Microwave Filter Company 17842-6 low-pass filter (passband 0-120 MHz, nominal 0.5 dB insertion loss, and minimum 50 dB rejection between 127 and 488.5 MHz) was used. Most significantly, a custom-built, inductively resonant tank circuit, i.e., an impedance-matching circuit, was developed to maximize the power transfer from the RF amplifier to the coil [S7], as shown in Figure S1(c). While previous EDMR experiments yielded $B_1 \sim 2$ mT for an electrical drive power of ~200 W [S1], this setup produced $B_1 \sim 5$ mT with ~4 W. This efficient power transfer from the tank circuit also implied limited reflections, resulting in the suppression of RF-induced, spin-independent electrical currents and further improving the detection of the spin-dependent electrical currents. The tank circuit also acted as a narrow bandpass filter that, in addition to the low-pass RF filter, removed higher RF harmonics as well as potential one-photon resonances superimposing upon the multi-photon resonances. Indirectly, the tank circuit also contributed to fewer amplifier harmonics by allowing the RF amplifier to be operated well within its linear response range.

B) "Copper box" shielding of the RF coil from the adjacent RF tank circuit

The copper box around the coil shown in Figure S1(a) was built to reduce the inhomogeneity of $B_1$, after we observed that the two-photon linewidths were much broader than those reported previously from SY-PPV OLEDs using a non-resonant coil and a different sample-holder setup [S1], as shown by the data in Figure S*2*. We attribute the distorted EDMR spectrum to $\boldsymbol{B_1}$ inhomogeneities, given that the RF coil was placed in the middle of the tank circuit components, as shown in Figure S1(a). Under strong drive, stray electromagnetic fields from the circuit elements may interfere with $\boldsymbol{B_1}$ generated by the coils. After adding the copper encasement, we again observed two-photon resonances resembling the previously measured line shapes [S1], although the one-photon resonances were not entirely identical. We conclude that the copper box changed the electromagnetic environment around the coil, even though it likely did not entirely diminish the overall inhomogeneity of $B_1$. Furthermore, it is plausible that the copper box changes the overall angle of $\boldsymbol{B_1}$ relative to $\boldsymbol{B_0}$, which could result in different intensity ratios between one- and two-photon resonances, as discussed in the main text in the context of the angular dependence of multi-photon EDMR lines. In either case, the addition of the copper box does not qualitatively affect the line shifts of multi-photon resonances or their mutual ratios, which are discussed in this study.

C) Determination of RF field-modulation frequency for lock-in detected cw EDMR spectroscopy

To determine an optimal modulation frequency for the lock-in detected cw EDMR experiments, we measured a modulation frequency dependence of the EDMR spectra of SY-PPV OLEDs at room temperature between 5 Hz and 20 kHz for low RF power, as shown in Figure S3(a). Figure S3(b) displays the intensity of the one-photon resonance aligned in the in-phase channel, as well as the phase adjustment needed to align the signal in the in-phase channel, showing that the lock-in detected signals represent steady-state rates at modulation frequencies at or below 20 Hz, i.e., the signal intensity begins to drop above 20 Hz. We found that, aside from signal intensities, neither low-RF power nor high-RF power EDMR spectra were significantly affected by modulation frequencies, as evidenced by Figure S4 showing EDMR obtained at 1 kHz modulation frequency. This measurement corresponds to the data shown in Figure 1(b) of the main text, obtained under similar drive-strength conditions under a modulation frequency of 20 Hz. While we chose to conduct most measurements at 20 Hz in order to satisfy the steady-state condition assumed in the theoretical model and to maximize the EDMR signal strength, some of the data recorded were measured at 1 kHz (as noted) because higher modulation frequencies permit much faster sweep rates and enable the separation of fast, drive-induced spin-independent currents from slower, spin-dependent recombination currents, following a previously described procedure [S1].

### D) Utilization and limitations of RF coil current monitor

The current through the 0.1 Ω monitoring resistor shown in Figure S1(c) and, thus, the voltage drop across it that can be detected using an oscilloscope, is proportional to the coil current and thus $B_1$. However, the actual proportionality factor between $B_1$ and the observed voltage changes with the tank circuit's tuning state and, thus, whenever a sample is changed or moved within the setup or when resistive heating changes the wire resistances and therefore the impedances within the tank circuit. Figure S5 illustrates the effect of a change in circuit tuning during a measurement series. It shows two spectra recorded on the same device at nearly the same current amplitude across the monitor, the black curve recorded during the sequence of increasing power, and the gray curve later during the sequence of decreasing power. Even though the spectra were recorded at the same nominal power, the greater amplitude of the inverted one-photon resonance (gray curve) indicates a significantly larger $B_1$ than for the black curve, revealing the circuit's changed tuning state. Moreover, the gray curve had a slightly lower RF monitor current, despite it having the higher $B_1$ field. These observations underscore the challenge of independently measuring $B_1$, and, thus, the importance of finding ways to test the theory without accurate knowledge of $B_1$, as is provided by Eq. (2) in the main text.

### E) Replots of the data in Figure 3 of the main text

The data shown in Figure 3 of the main text are replotted in Figure S6 on scales adjusted to match the theoretically predicted ratios from Eq. (2) of the main text to diagonal lines.

### F) EDMR spectra under non-perpendicular orientation of $B_1$ and $B_0$

Three measurement series, each with a different OLED, were conducted with non-perpendicular orientation of $\boldsymbol{B_1}$ and $\boldsymbol{B_0}$, i.e., with the angle $\theta$ between $\boldsymbol{B_1}$ and $\boldsymbol{B_0}$ set to 69°, the physical limit of the sample holder. In the first series, we raised $B_1$ with the sample holder in its usual position (i.e., $\boldsymbol{B_1} \perp \boldsymbol{B_0}$) and then decreased it again with the angle changed ($\theta = 69°$). In the second series, we raised and decreased the drive power with the sample holder turned to $\theta = 69°$ and then raised and decreased the power again with $\boldsymbol{B_1} \perp \boldsymbol{B_0}$. In the third series, we raised $B_1$ with the sample holder at $\theta = 69°$ and then decreased it again with $\boldsymbol{B_1} \perp \boldsymbol{B_0}$.

Figure S7 displays a sample of spectra with $\theta = 69°$ from one of the measurement series. As predicted by Eq. (3) of the main text, a strong two-photon resonance is seen, yet no discernable three-photon resonance.

### G) Numerical simulations of EDMR spectra under non-perpendicular orientation of $B_1$ and $B_0$

Figure S8 shows three-photon resonance line centers obtained from numerical simulations for $\theta = 65°$ to 90° in intervals of 6°. The simulations apply the Floquet approach to solve the density matrix Liouville equation [S8]. The peak centers from the simulations were determined from the peak extrema as described in Ref. [10]. Fitting the peak centers to a first-order polynomial $y = ax + b$, where $x = |\sin\theta - (9/8)\sin^3\theta|$, shows excellent agreement with the angular dependence expected from Eq. (3) of the main text, thus confirming the validity of another way to scrutinize the three-photon resonances experimentally.

## H) Breakdown of all data used for the study presented

Figure S9 through Figure S14 display all data obtained in the course of 1959 EDMR measurements. Figure S9, Figure S10, and Figure S11 show the resulting averages of one-, two-, and three-photon resonance peak centers, respectively, as a function of the RF monitor current for each OLED studied. Figure S12 plots the shifts of three- vs. two-photon resonance lines for the averaged peak centers of each device. Samples 7-12 were measured with the copper box in place. The open symbols fall outside of the strong-drive limit defined in the main text. The light-blue points were measured under 1 kHz modulation. The red points were measured at an angle of 69°. The solid line passes through the origin and has a slope equal to the ratio ($\Delta B_{\gamma_3}/\Delta B_{\gamma_2} = 9/16$) predicted by Eq. (2) of the main text. Figure S13 and Figure S14 are, respectively, similar plots of shifts of three- vs. one-photon and two- vs. one-photon resonance lines for the averaged peak centers of each device.

**Supplemental Figures**

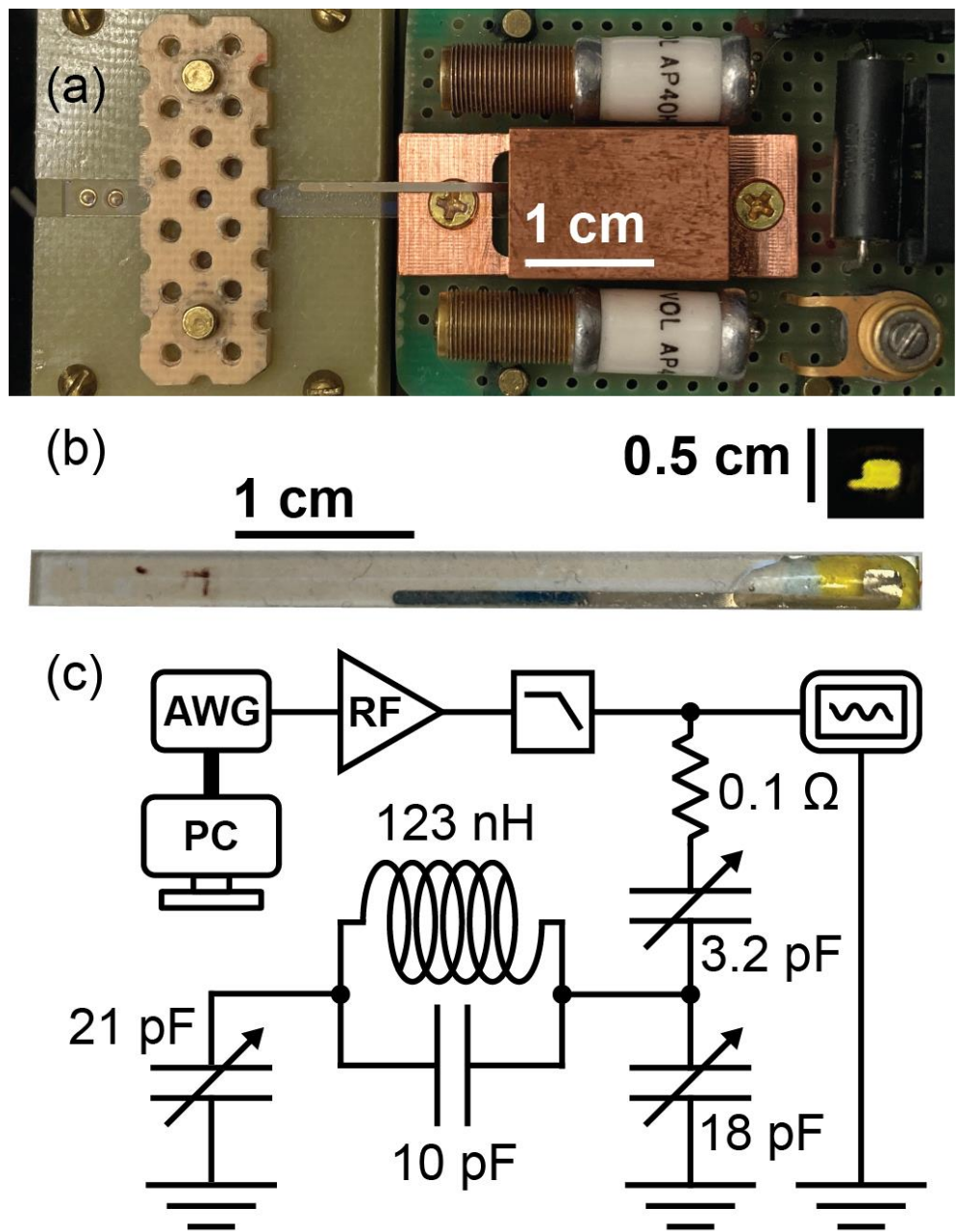

Figure S1. EDMR setup built for the high $B_1$ experiments. (a) Photo of the sample holder, connecting an SY-PPV OLED device with 2 mm × 3 mm rectangular active area that was deposited at the tip of the EDMR contact template, located inside the RF coil. The RF coil itself is placed inside the copper box shown in the photo. Components of the *tank* circuit are fixed around the copper box. (b) Photo of the EDMR contact template with an OLED sample on the right side, together with a view of the OLED pixel emitting under bias. (c) RF circuit diagram creating an impedance matching network, i.e. an inductively resonant *tank* circuit that was used for the experiments presented.

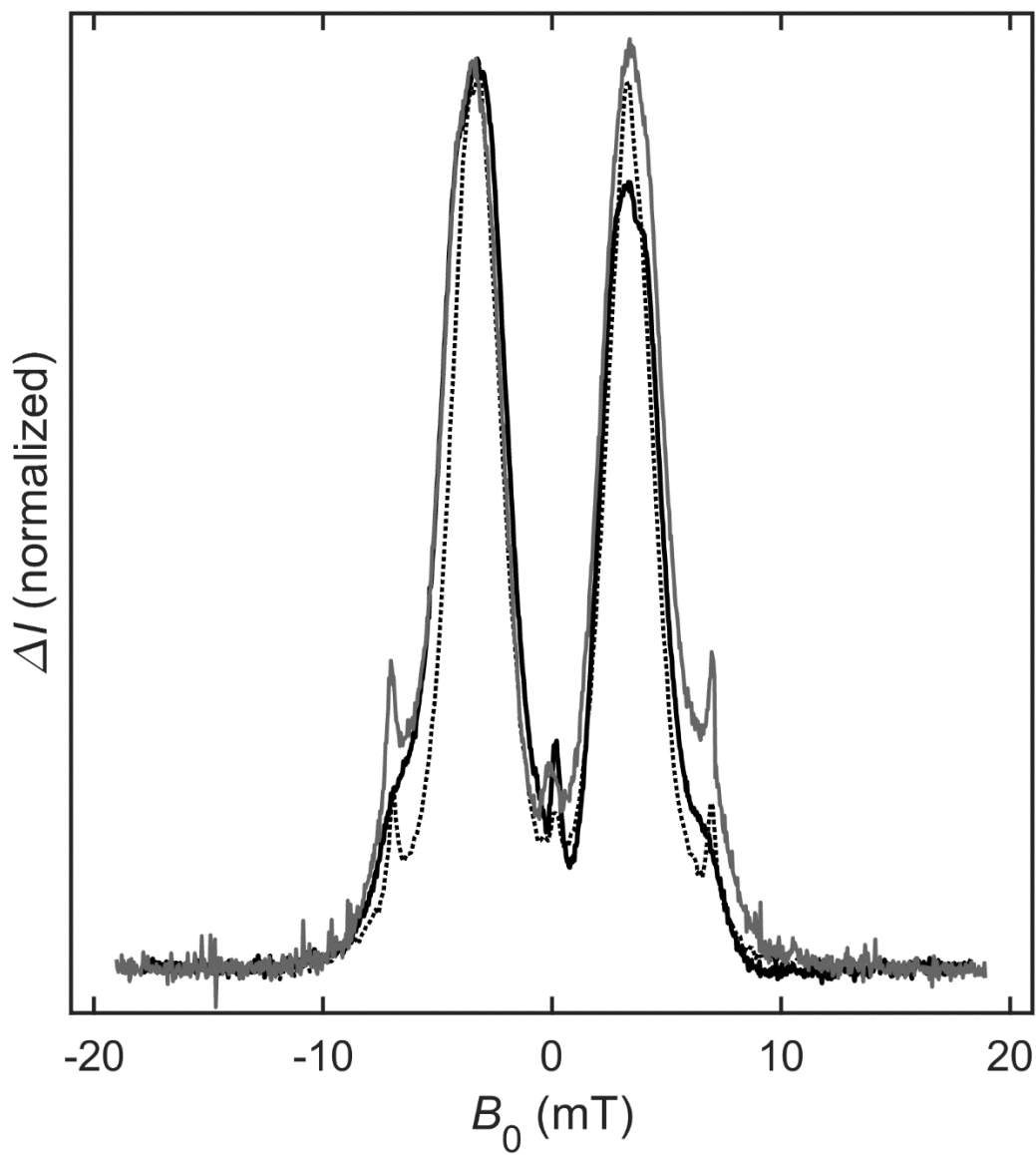

Figure S2. Three EDMR spectra obtained from identically prepared OLEDs, measured i) in a previously reported non-resonant EDMR coil setup [S1] (dashed); ii) with the tank-circuit coil described in the main text, yet without copper box (solid black); and iii) with the tank-circuit coil with the copper box (gray). All three data sets display two-photon resonances. However, in the absence of the copper box, the spectrum of the tank circuit displays strong distortions.

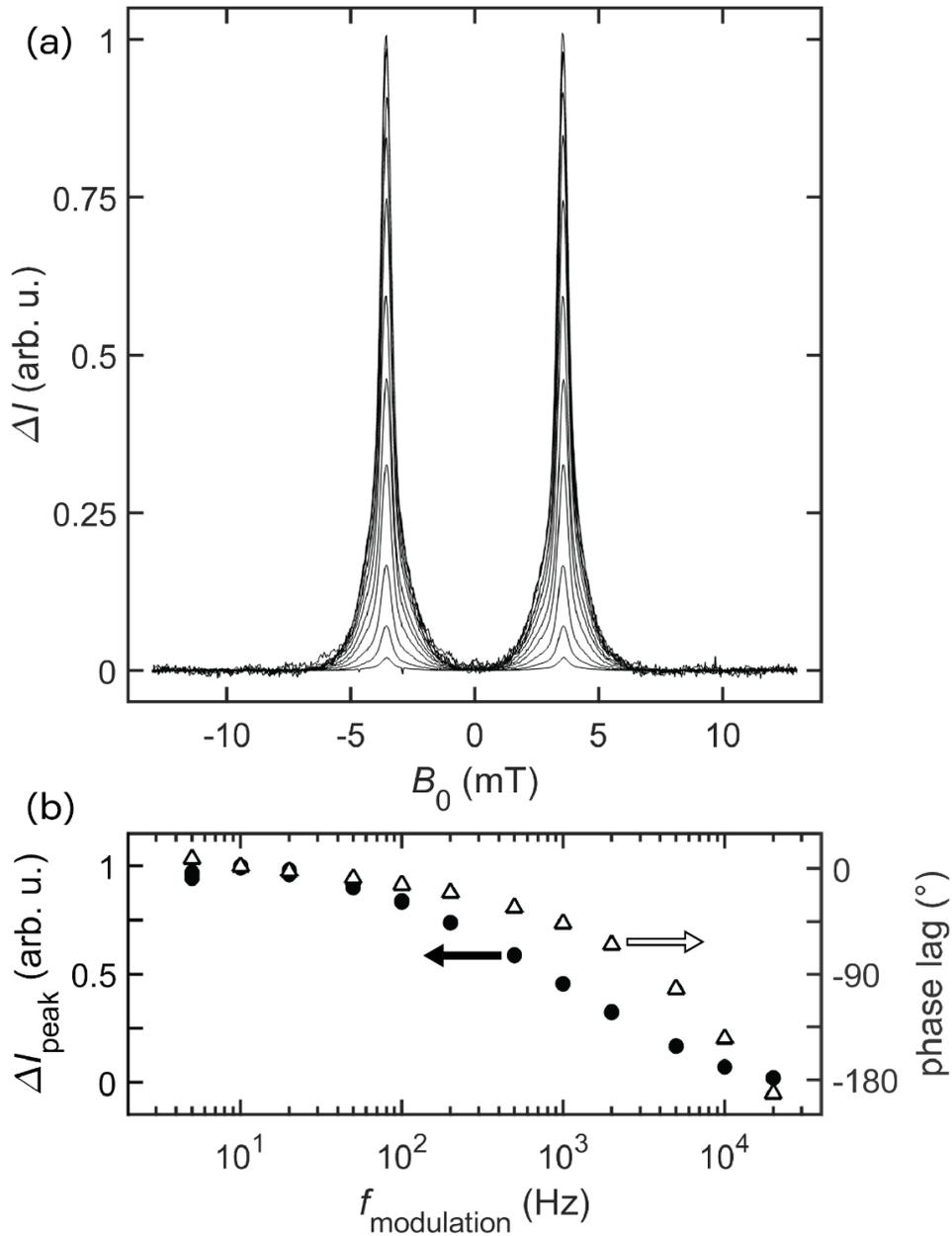

Figure S3. (a) Plot of several lock-in detected, RF field-modulated cw EDMR spectra, measured at low and constant power with varying modulation frequency. The reference phase of the lock-in amplifier has been adjusted post-measurement so that all of the spin-dependent resonance signal is aligned in the in-phase channel. (b) *Left-hand axis*: intensity of the one-photon peak centers from (a) as a function of modulation frequency. The intensity begins to decrease above 20 Hz. The data include points that are not in the selection shown in (a). *Right-hand axis*: phase lag of the spectra in (a) as a function of modulation frequency. The phase lag is equivalent to the reference phase required to align all of the spin-dependent resonance in the in-phase channel.

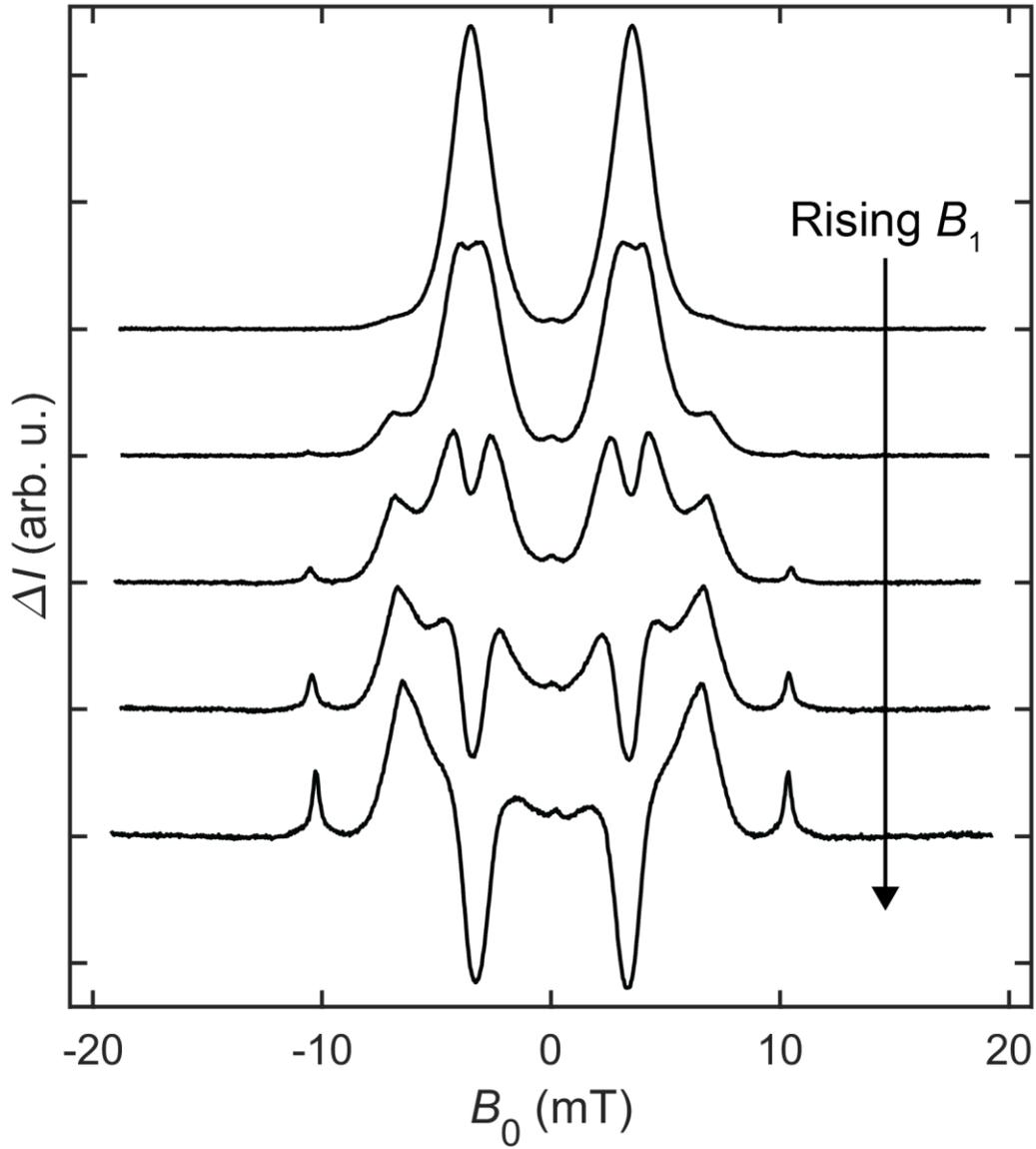

Figure S4. A selection of EDMR spectra measured under similar drive-strength conditions as the data shown in Figure 1 of the main text, with 1 kHz modulation instead of 20 Hz modulation of the RF drive. No qualitative differences are seen, nor quantitative differences with regard to the resonance centers (see Figure S9-Figure S14).

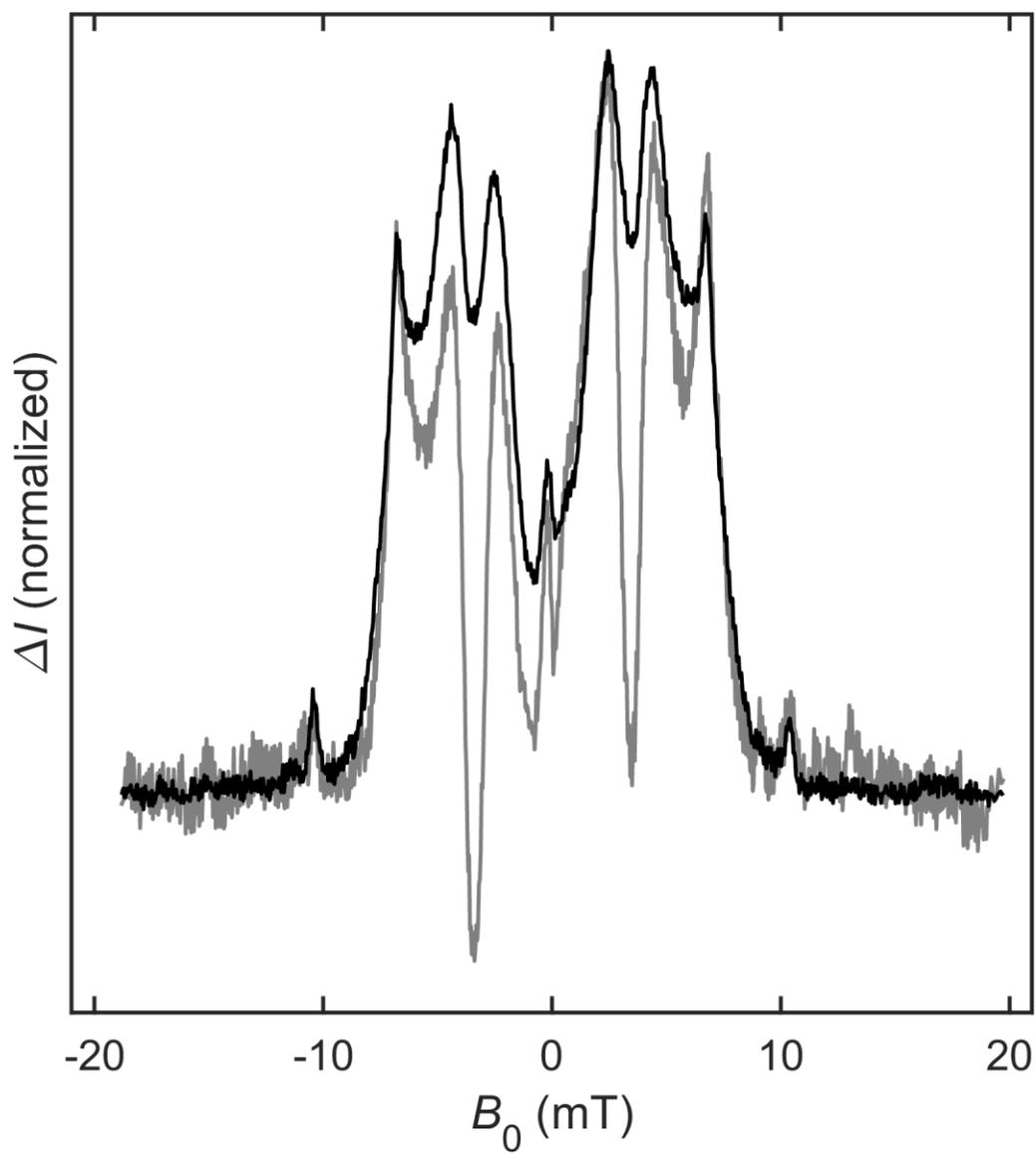

Figure S5. EDMR spectra measured at the same nominal drive power during the sequence of ascending (black) and descending (gray) RF power. The different depths of the one-photon peak inversion indicate that the two spectra were measured under different $B_1$, even though the input power was the same.

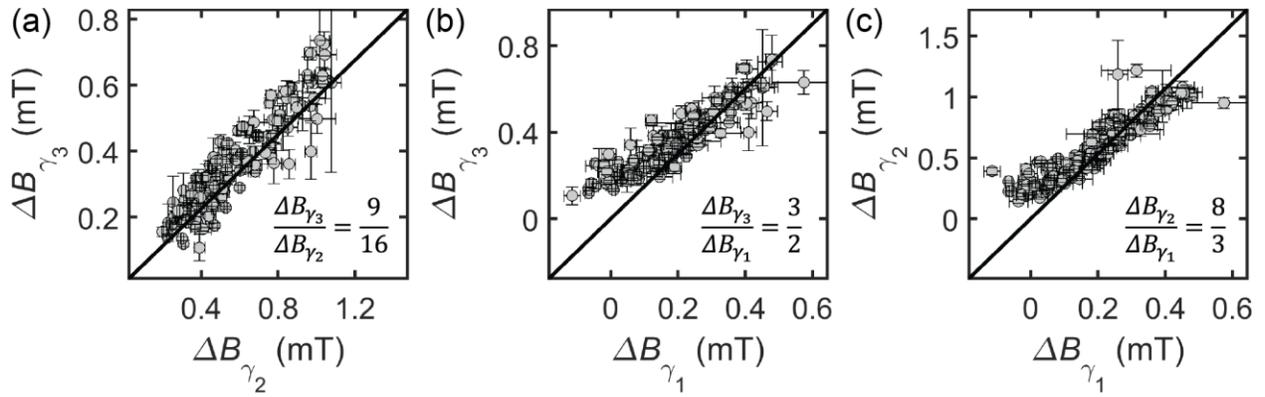

Figure S6. Shifts of (a) three- vs. two-photon, (b) three- vs. one-photon, (c) and two- vs. one-photon resonance line centers of all OLED samples that fall within the strong-drive limit defined in the main text. The solid lines pass through the origin and have slopes equal to the ratio shown in each panel. The ratios are predicted by Eq. (2) of the main text.

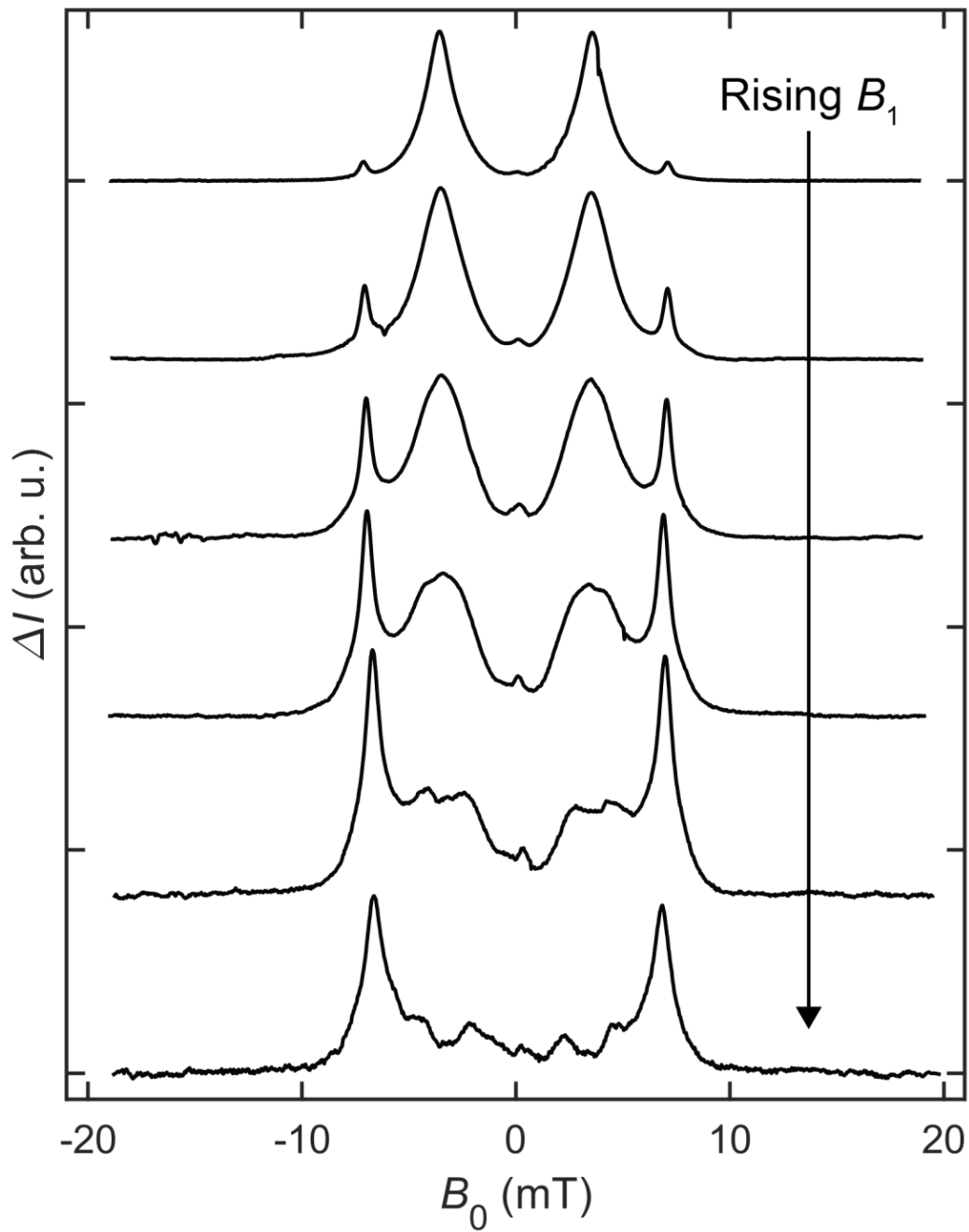

Figure S7. Selection of EDMR spectra with a nominal angle of 69° between $B_1$ and $B_0$ rather than 90°, measured under similar drive-strength conditions to the data in Figure 1(b) of the main text. The RF input power increases from top to bottom.

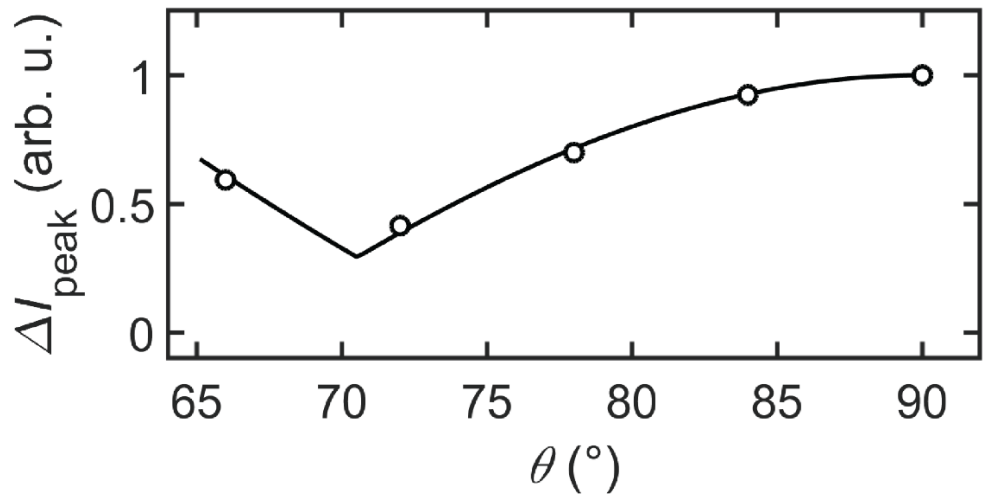

Figure S8. Resonance peak intensities (open circles) obtained from numerical simulations following the Liouville density matrix approach, showing consistency with the analytical Eq. (3) of the main text (solid curve).

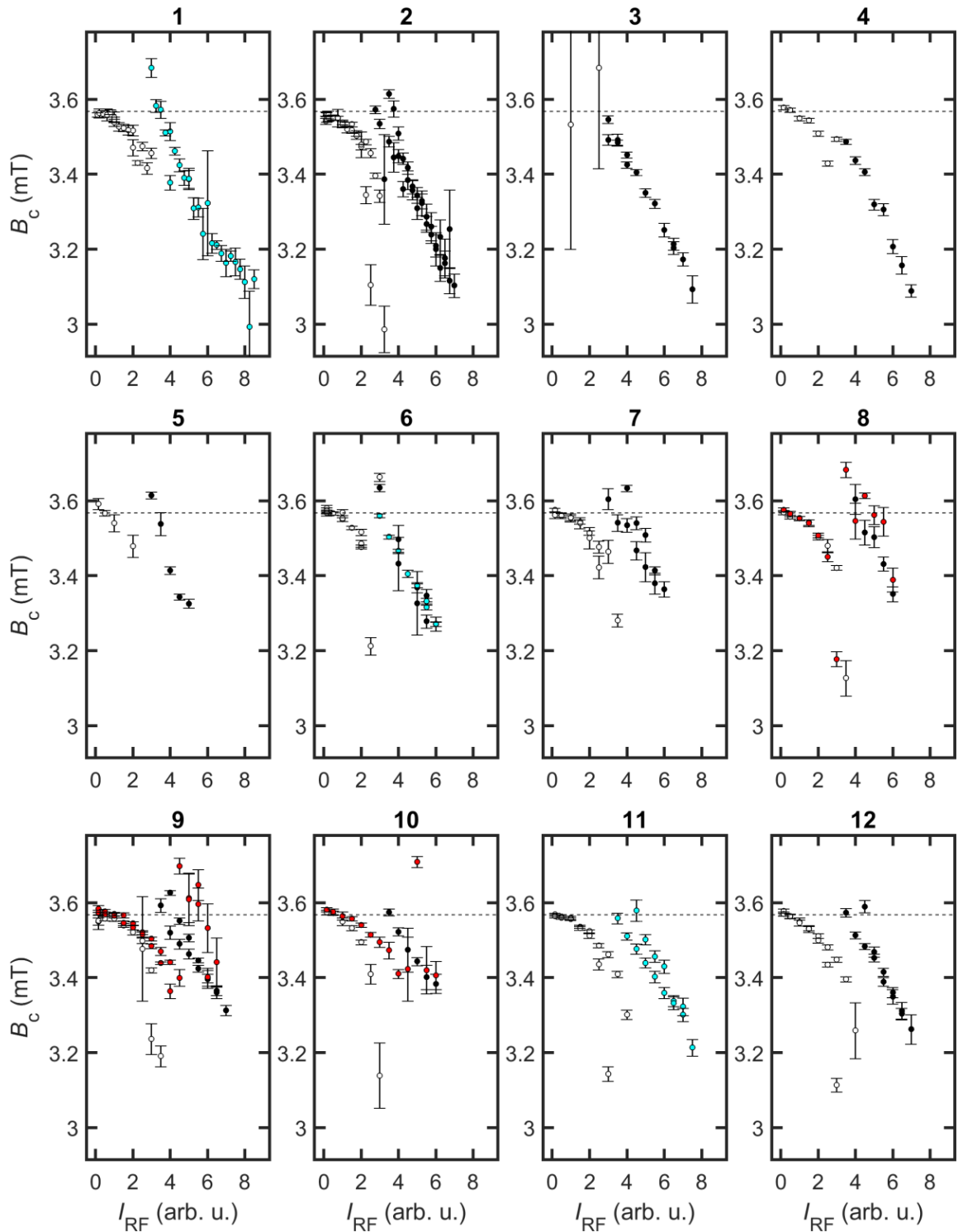

Figure S9. Averages of one-photon resonance peak centers vs. RF monitor current for each individual OLED sample. Samples 7-12 were measured with the copper box. The open symbols fall outside of the strong-drive limit defined in the main text. The light-blue points were measured under 1 kHz modulation. The red points were measured at an angle of $\theta = 69°$. The horizontal dashed line indicates the unshifted resonance.

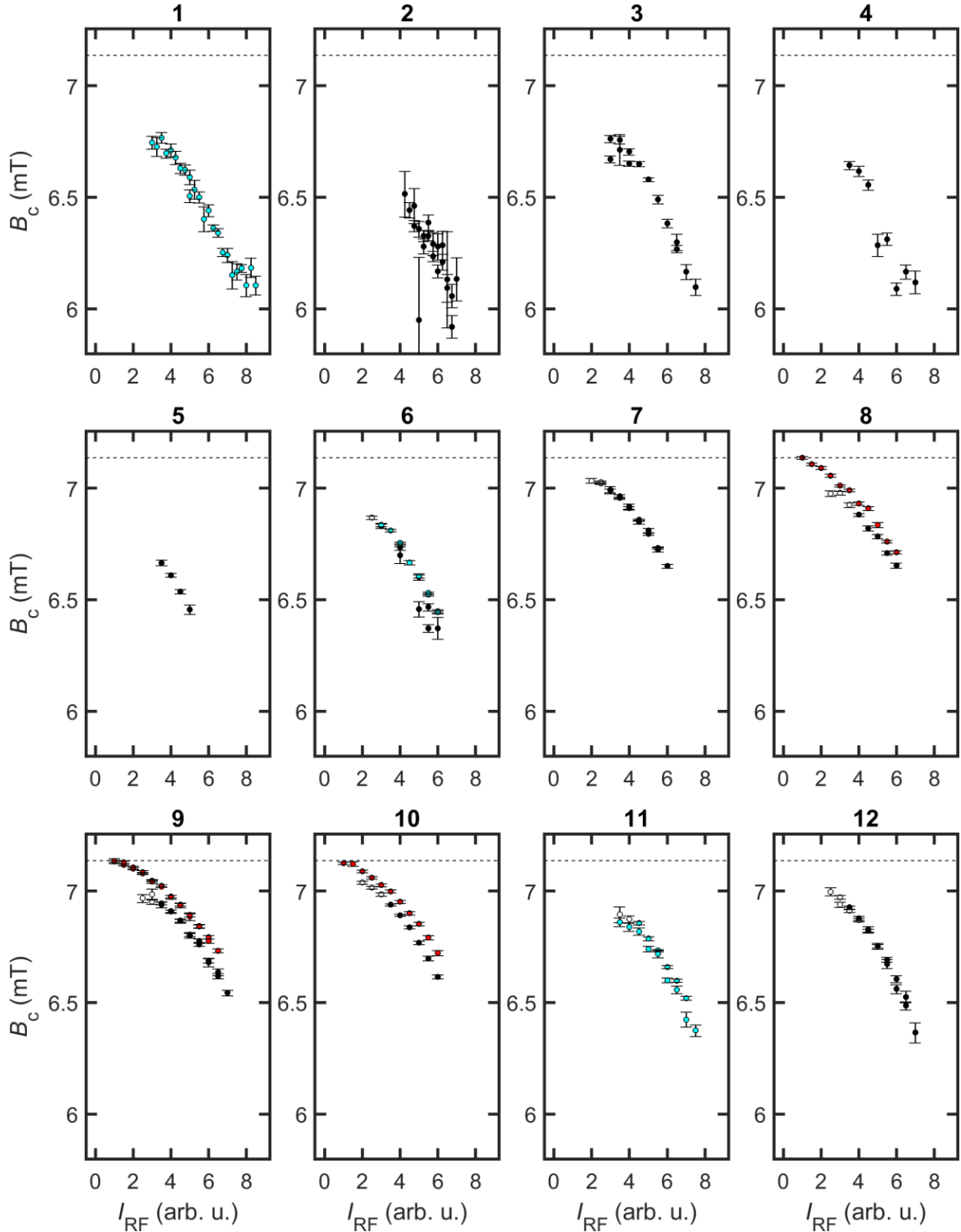

Figure S10. Averages of two-photon resonance peak centers vs. RF monitor current for each individual OLED sample. Samples 7-12 were measured with the copper box. The open symbols fall outside of the strong-drive limit defined in the main text. The light-blue points were measured under 1 kHz modulation. The red points were measured at an angle of $\theta = 69°$. The horizontal dashed line indicates the unshifted resonance.

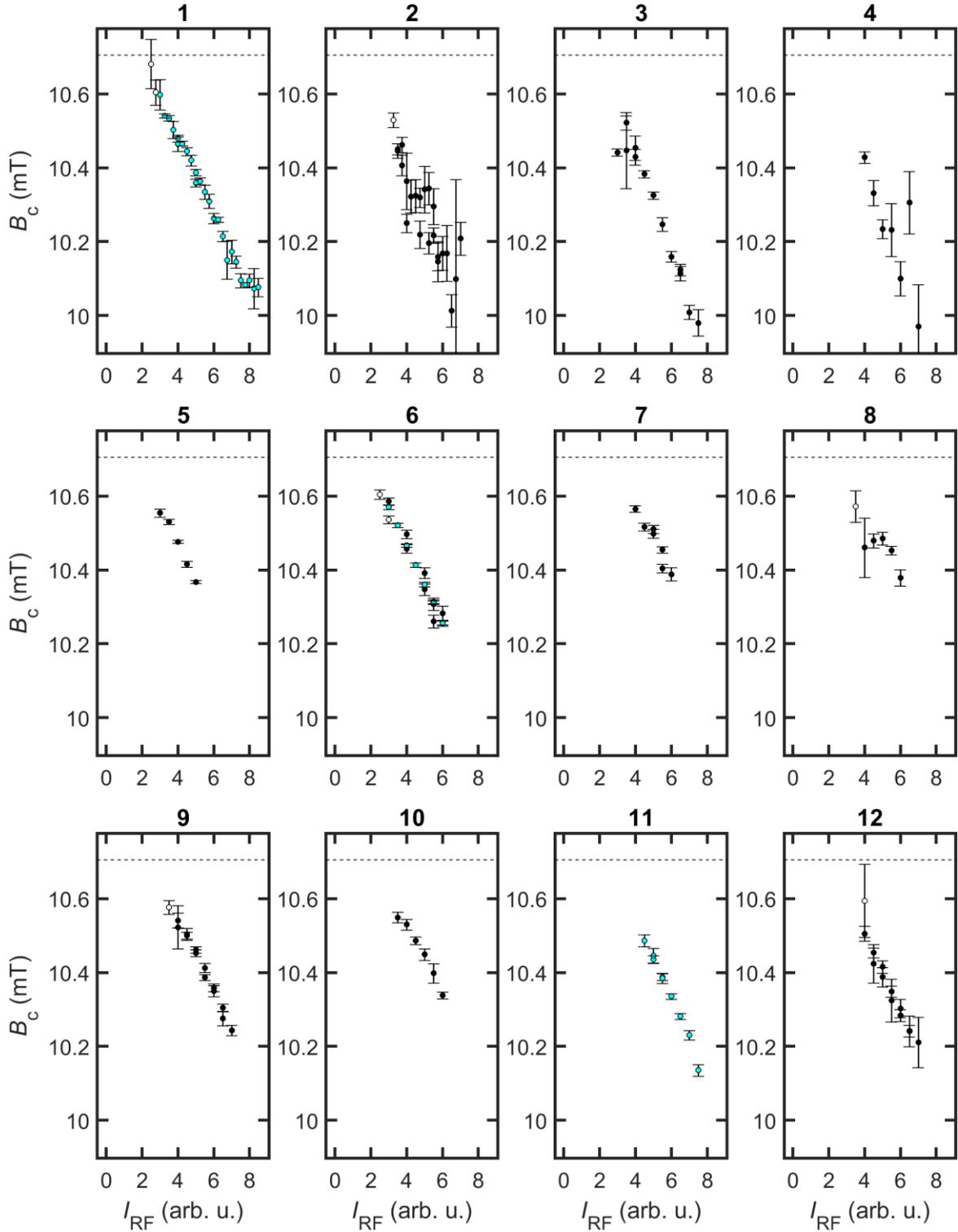

Figure S11. Averages of three-photon resonance peak centers vs. RF monitor current for each individual OLED sample. Samples 7-12 were measured with the copper box. The open symbols fall outside of the strong-drive limit defined in the main text. The light-blue points were measured under 1 kHz modulation. The red points were measured at an angle of $\theta = 69°$. The horizontal dashed line indicates the unshifted resonance.

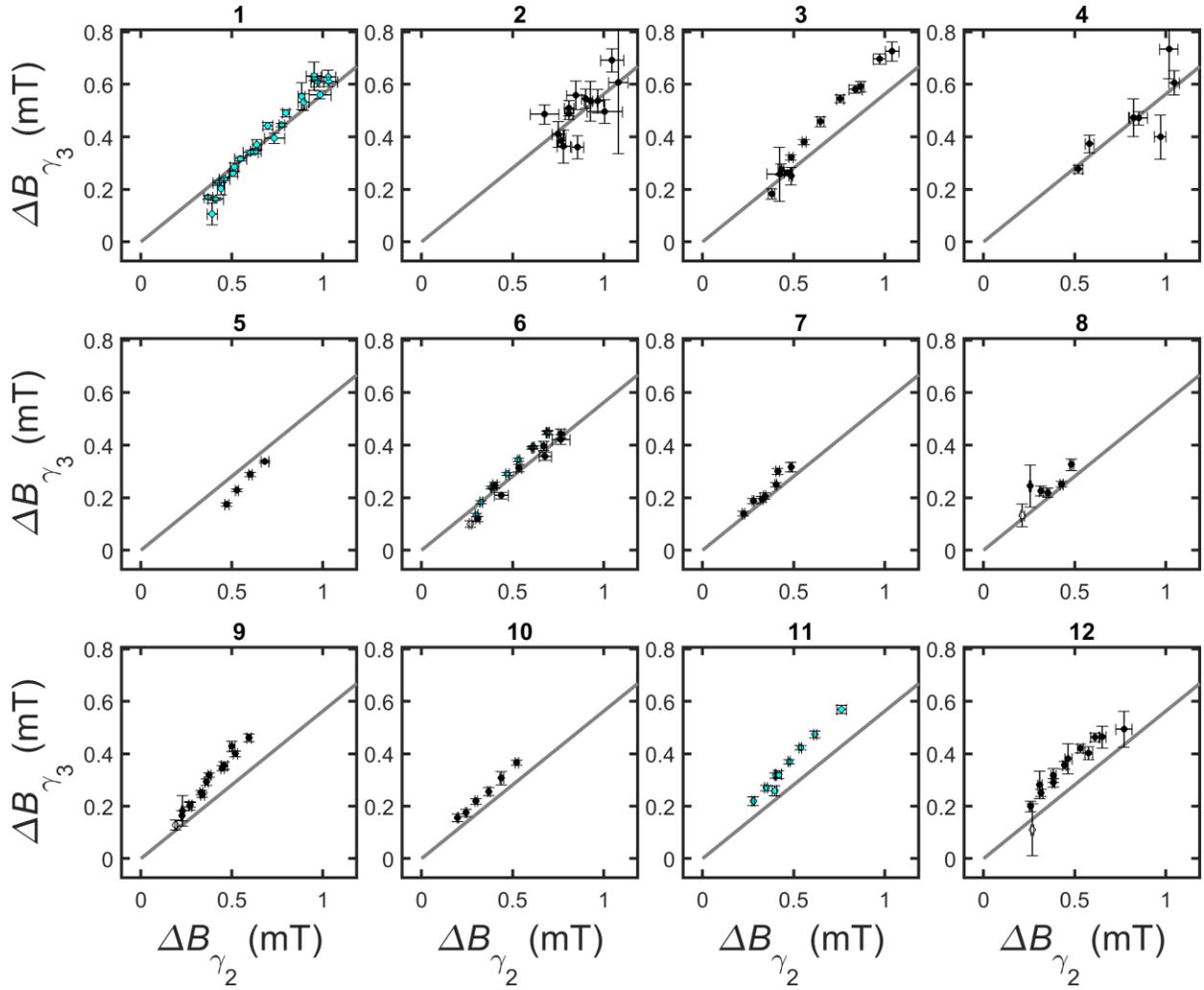

Figure S12. Shifts of three- vs. two-photon resonances for the averaged peak centers of each OLED sample. Samples 7-12 were measured with the copper box. The open symbols fall outside of the strong-drive limit defined in the main text. The light-blue points were measured under 1 kHz modulation. The red points were measured at an angle of $\theta = 69°$. The solid line passes through the origin and has slope equal to the ratio ($\Delta B_{\gamma_3}/\Delta B_{\gamma_2}= 9/16$) predicted by Eq. (2) in the main text.

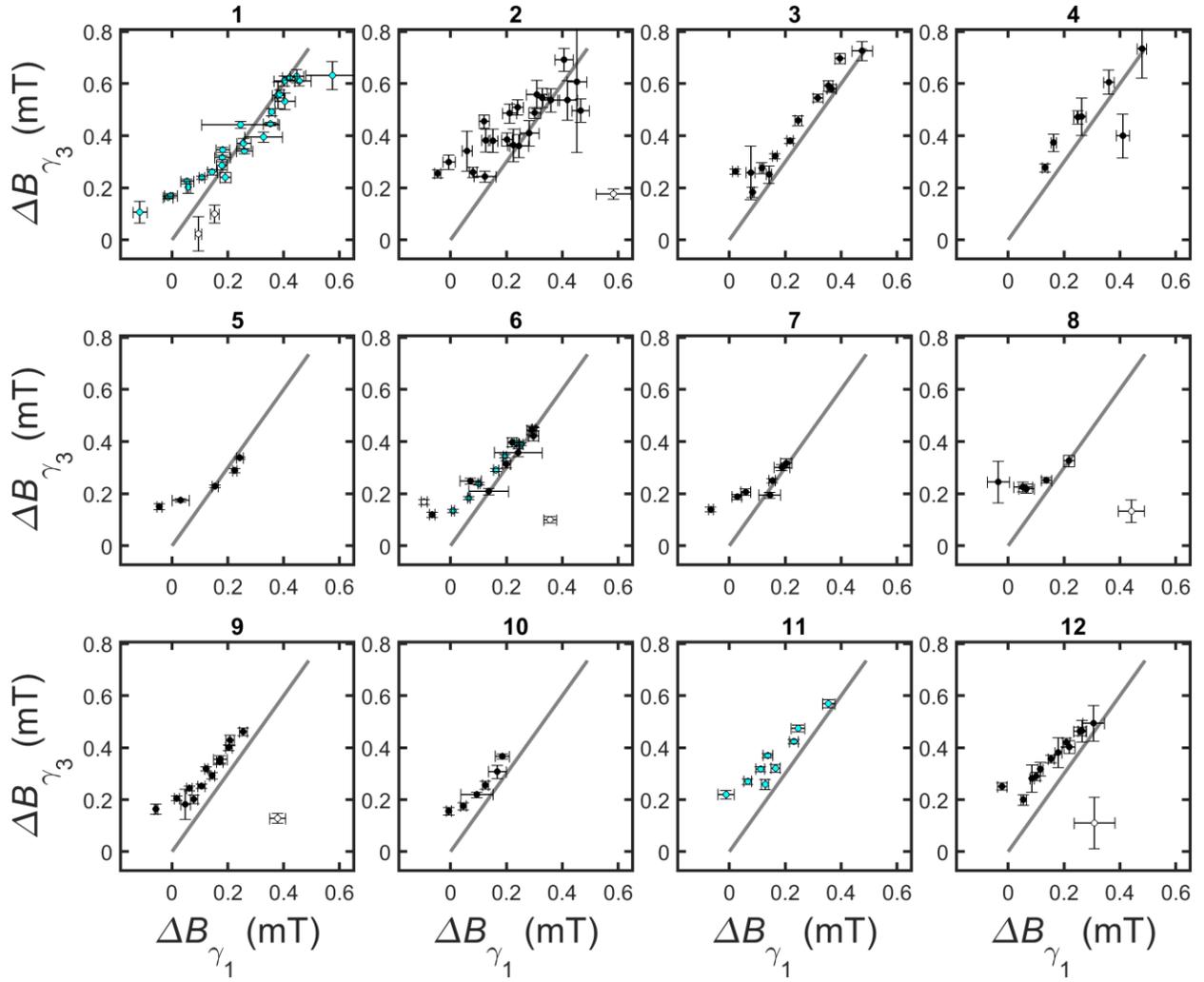

Figure S13. Shifts of three- vs. one-photon resonances for the averaged peak centers of each OLED sample. Samples 7-12 were measured with the copper box. The open symbols fall outside of the strong-drive limit defined in the main text. The light blue points were measured under 1 kHz modulation. The red points were measured at an angle of $\theta = 69°$. The solid line passes through the origin and has a slope equal to the ratio ($\Delta B_{\gamma_3}/\Delta B_{\gamma_1}= 3/2$) predicted by Eq. (2) in the main text.

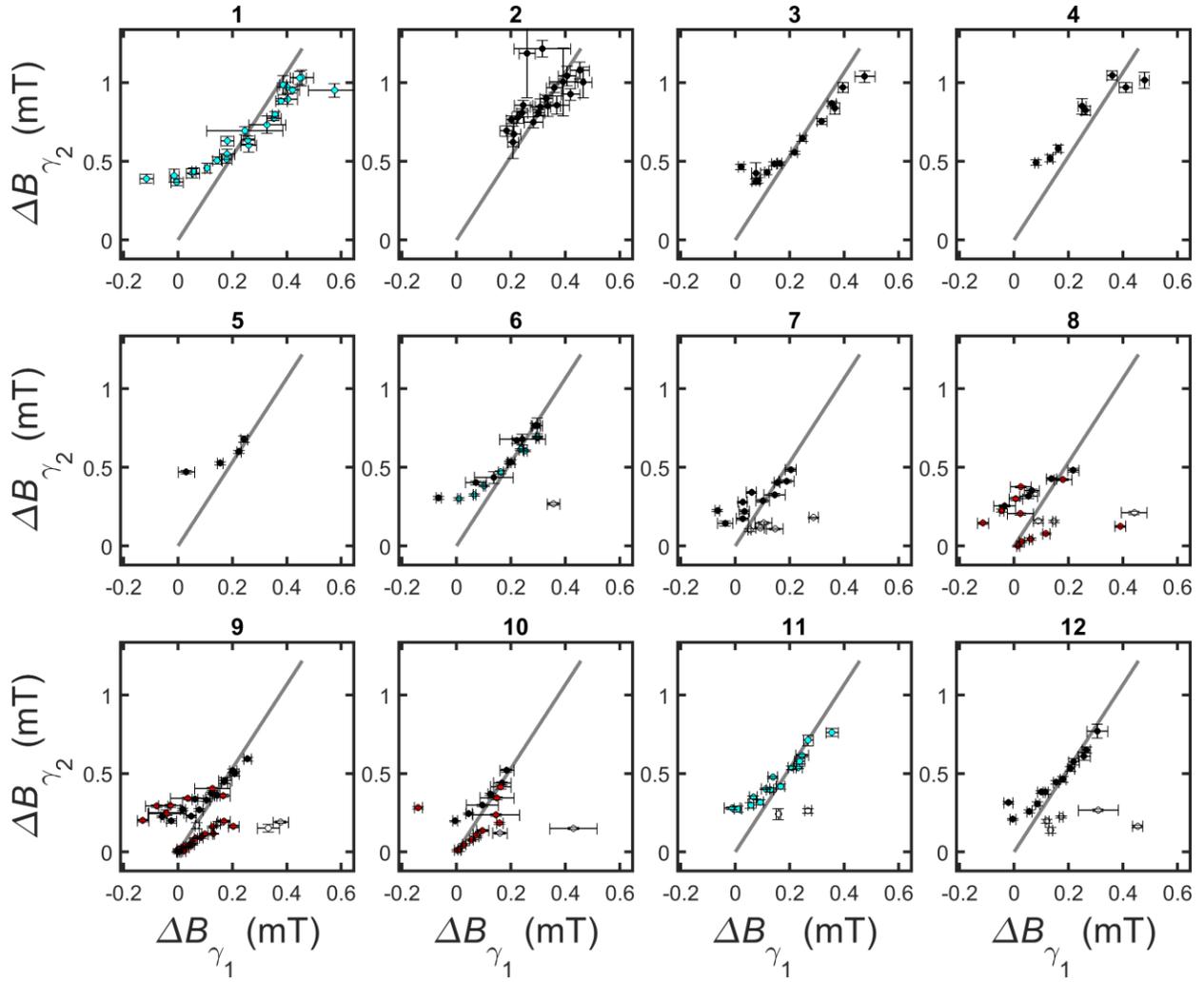

Figure S14. Shifts of two- vs. one-photon resonances for the averaged peak centers of each OLED sample. Samples 7-12 were measured with the copper box. The open symbols fall outside of the strong-drive limit defined in the main text. The light-blue points were measured under 1 kHz. The red points were measured at an angle of $\theta = 69°$. The solid line passes through the origin and has a slope equal to the ratio ($\Delta B_{\gamma_2}/\Delta B_{\gamma_1} = 8/3$) predicted by Eq. (2) in the main text.

**Supplemental References**

S1. S. I. Atwood, S. Hosseinzadeh, V. V. Mkhitaryan, T. H. Tennahewa, H. Malissa, W. Jiang, T. A. Darwish, P. L. Burn, J. M. Lupton, C. Boehme. arXiv:2310.14180 (cond-mat.mes-hall) (2023). Under review at *Phys. Rev. B*. https://doi.org/10.48550/arXiv.2310.14180

S2. D. P. Waters, G. Joshi, M. Kavand, M. E. Limes, H. Malissa, P. L. Burn, J. M. Lupton, and C. Boehme, *Nat. Phys.* 11, 910 (2015). https://doi.org/10.1038/nphys3453

S3. D. R. McCamey, H. A. Seipel, S.-Y. Paik, M. J. Walter, N. J. Borys, J. M. Lupton, and C. Boehme. *Nat. Mater.* 7, 723 (2008). https://doi.org/10.1038/nmat2252

S4. M. Kavand, D. Baird, K. van Schooten, H. Malissa, J. M. Lupton, and C. Boehme. *Phys. Rev. B* 94, 075209 (2016). https://doi.org/10.1103/PhysRevB.94.075209

S5. H. Malissa, M. Kavand, D. P. Waters, K. J. van Schooten, P. L. Burn, Z. V. Vardeny, B. Saam, J. M. Lupton, and C. Boehme. *Science* 345, 1487 (2014). https://doi.org/10.1126/science.1255624

S6. T. H. Tennahewa, S. Hosseinzadeh, S. I. Atwood, H. Popli, H. Malissa, J. M. Lupton, C. Boehme. *Phys. Rev. B* (in print). https://doi.org/10.48550/arXiv.2207.07086

S7. J. Mispelter, M. Lupu, and A. Briguet. *NMR probeheads for biophysical and biomedical experiments.* Imperial College Press: London (2015).

S8. S. Jamali, V. V. Mkhitaryan, H. Malissa, A. Nahlawi, H. Popli, T. Grünbaum, S. Bange, S. Milster, D. M. Stoltzfus, A. E. Leung, et al., *Nat. Commun.* 12, 465 (2021). https://doi.org/10.1038/s41467-020-20148-6